\documentclass[pdflatex,sn-mathphys-num]{sn-jnl}% Math and Physical Sciences Numbered Reference Style
\usepackage{graphicx}%
\usepackage{multirow}%
\usepackage{amsmath,amssymb,amsfonts}%
\usepackage{amsthm}%
\usepackage{mathrsfs}%
\usepackage[title]{appendix}%
\usepackage{xcolor}%
\usepackage{textcomp}%
\usepackage{manyfoot}%
\usepackage{booktabs}%
\usepackage{algorithm}%
\usepackage{algorithmicx}%
\usepackage{algpseudocode}%
\usepackage{listings}%

\usepackage[numbers]{natbib}
\usepackage{float}
\usepackage{pgfplots}
\usepackage{subcaption}
\usepackage{tikz}
\usepackage{placeins}
\pgfplotsset{compat=1.18}

\definecolor{clrRandom}{RGB}{0,0,0}
\definecolor{clrTabu}{RGB}{31,119,180}
\definecolor{clrSim}{RGB}{255,127,14}
\definecolor{clrHybrid}{RGB}{44,160,44}
\definecolor{clrKpp}{RGB}{214,39,40}

\theoremstyle{thmstyleone}%
\theoremstyle{thmstyletwo}%

\theoremstyle{thmstylethree}%

\begin{document}

\title[Article Title]{Adaptive Quantum Optimized Centroid Initialization}


\author*[1]{\fnm{Nicholas R.} \sur{Allgood}}\email{allgood1@umbc.edu}

\author[1]{\fnm{Ajinkya} \sur{Borle}}\email{aborle1@umbc.edu}

\author[1]{\fnm{Charles K.} \sur{Nicholas}}\email{nicholas@umbc.edu}

\affil*[1]{\orgdiv{Department of Computer Science and Electrical Engineering}, \orgname{University of Maryland Baltimore County}, \orgaddress{\street{1000 Hilltop Circle}, \city{Baltimore}, \postcode{21250}, \state{MD}, \country{US}}}

%\affil[2]{\orgdiv{Department}, \orgname{Organization}, \orgaddress{\street{Street}, \city{City}, \postcode{10587}, \state{State}, \country{Country}}}

%\affil[3]{\orgdiv{Department}, \orgname{Organization}, \orgaddress{\street{Street}, \city{City}, \postcode{610101}, \state{State}, \country{Country}}}


\abstract{
Prototype-based clustering algorithms such as k-means are sensitive to the selection of initial cluster centroids, with poor initialization leading to slower convergence and suboptimal solutions trapped in local minima. We present Adaptive Quantum Optimized Centroid Initialization (AQOCI), a method that formulates the centroid initialization problem as a Quadratic Unconstrained Binary Optimization (QUBO) problem and solves it using quantum annealing or quantum-inspired solvers. AQOCI extends a prior method (QOCI) by introducing an iterative refinement mechanism inspired by the Gauss-Seidel and Jacobi methods, enabling the recovery of real-valued centroid coordinates from binary solver outputs through adaptive scaling and offset adjustments.
We evaluate AQOCI using three solver backends:  TABU search, simulated annealing, and D-Wave's HybridBQM on synthetic Gaussian data with controlled sweeps over cluster separation, cluster count, dimensionality, and sample size, as well as on the MOTIF malware classification dataset, comparing against standard k-means with random initialization and k-means++ initialization. On the MOTIF dataset, AQOCI produces clusterings that are competitive with and, at smaller sample sizes, superior to k-means++, with V-measure improvements of up to 26\%. On synthetic data with heavily overlapping clusters, AQOCI--SimAnn outperforms k-means++ in V-measure. On well-separated synthetic data, k-means++ is clearly superior, and AQOCI exhibits a consistent performance plateau attributable to the binary encoding resolution. The dimensionality sweep demonstrates scalability to at least $d = 10$ without degradation. These results demonstrate that QUBO-based centroid initialization is a viable alternative formulation whose performance advantage is data-structure-dependent, strongest when clusters overlap and sample sizes are small, and whose potential for quantum speedup may be realized as quantum annealing hardware continues to mature.
}




\maketitle

\section{Introduction}

Cluster analysis is fundamental to unsupervised machine learning, with
applications spanning classification, anomaly detection, and data
compression. Among clustering methods, prototype-based algorithms such as
$k$-means \cite{lloyd_clustering, MacQueen1967} remain widely used due to their
simplicity and scalability. However, $k$-means is well known to be sensitive
to the choice of initial centroids: different initializations can lead to
substantially different clusterings, and the algorithm provides no guarantee
of finding a global optimum \cite{MacQueen1967}.

The most widely adopted solution to this initialization problem is
$k$-means++ \cite{Arthur2007}, which selects initial centroids with
probability proportional to their squared distance from existing centers. This
approach has a provable $O(\log k)$ approximation guarantee and has become the
default initialization method in standard machine learning libraries. Despite
its effectiveness, $k$-means++ remains a sequential, greedy heuristic whose
performance can degrade on data with complex or overlapping cluster structure.

An alternative approach is to formulate centroid initialization as a
combinatorial optimization problem and solve it using quantum annealing. In
prior work, the authors introduced Quantum Optimized Centroid Initialization
(QOCI) \cite{Allgood2023}, which maps the initialization problem to a
Quadratic Unconstrained Binary Optimization (QUBO) formulation solvable on
quantum annealing hardware. QOCI demonstrated the feasibility of this
approach but was limited to integer-valued centroids in low-dimensional
settings and could not accommodate larger sample sizes within the constraints
of available quantum annealing resources.

In this paper, we present Adaptive Quantum Optimized Centroid Initialization
(AQOCI), which addresses these limitations through an iterative refinement
mechanism inspired by the Gauss-Seidel and Jacobi methods for solving linear
systems \cite{Pollachini2021}. AQOCI introduces adaptive scaling and offset
parameters that enable the recovery of real-valued centroid coordinates from
binary solver outputs, and decomposes the optimization into smaller blocks
that can be processed iteratively within the time constraints of quantum
annealing platforms.

We evaluate AQOCI on two datasets: synthetic Gaussian blobs and the MOTIF malware dataset \cite{Joyce2023}, using three solver backends (TABU search, simulated annealing, and D-Wave's Hybrid BQM) on the MOTIF data, and TABU search and simulated annealing on the expanded synthetic experiments. Critically, we compare against both random initialization and k-means++, providing a more rigorous baseline than previous evaluations. Our results show that the relative performance of
AQOCI depends on the structure of the data: on well-separated synthetic
clusters, $k$-means++ dominates all other methods, while on real-world malware
data with more complex structure, AQOCI produces competitive and sometimes
superior initializations, particularly at smaller sample sizes. We discuss
these results in the context of the formulation's potential for quantum
advantage as annealing hardware improves.

\section{Background}

\subsection{$k$-Means Clustering}

The $k$-means algorithm \cite{lloyd_clustering, MacQueen1967} partitions $n$
observations into $k$ clusters by minimizing the within-cluster sum of squared
distances:
\begin{equation}
    \min_S \sum_{i=1}^{k} \sum_{x \in S_i} \|x - \mu_i\|^2
\end{equation}
where $\mu_i$ is the mean of cluster $S_i$. The algorithm alternates between
assigning each observation to the cluster with the nearest centroid and
recomputing centroids as cluster means until convergence. While simple and
widely used, $k$-means provides no guarantee of finding a global optimum; the
final clustering depends on the choice of initial centroids, and the algorithm
is susceptible to convergence to local minima \cite{MacQueen1967}.

\subsection{$k$-Means++ Initialization}

Arthur and Vassilvitskii \cite{Arthur2007} introduced $k$-means++, a
randomized initialization procedure that selects centroids sequentially: the
first centroid is chosen uniformly at random from the data, and each
subsequent centroid is selected with probability proportional to the squared
distance from the nearest existing centroid. This biases selection toward
points that are far from existing centers, encouraging a spread of initial
centroids across the data. The method achieves an $O(\log k)$ approximation
guarantee in expectation and has become the default initialization in standard
implementations including scikit-learn. We include $k$-means++ as a baseline
in our experiments.

\subsection{Quantum Annealing and QUBO}

Quantum annealing exploits the adiabatic theorem to solve optimization
problems encoded as Hamiltonians. A system is initialized in the ground state
of a simple Hamiltonian and evolved slowly toward a problem Hamiltonian $H_p$;
if the evolution is sufficiently slow, the system remains near the
instantaneous ground state, yielding an approximate solution to the
optimization problem encoded in $H_p$.

In practice, problems are formulated as Quadratic Unconstrained Binary
Optimization (QUBO) instances, which are equivalent to Ising Hamiltonians. A
QUBO problem seeks a binary vector $x^* \in \{0,1\}^n$ minimizing
\begin{equation}
    f_Q(x) = x^T Q x = \sum_{i=1}^{n} \sum_{j=1}^{n} Q_{ij} x_i x_j
\end{equation}
where $Q \in \mathbb{R}^{n \times n}$ is an upper triangular weight matrix.
QUBO formulations are natively supported by D-Wave quantum annealers and can
also be solved by classical heuristics including TABU search and simulated
annealing, enabling benchmarking across solver backends. A comprehensive treatment of QUBO formulations and their applications across optimization domains is given by Glover et al.~\cite{Glover2019}.

Several prior works have explored quantum and QUBO-based approaches to clustering. Kumar et al.~\cite{Kumar2018} formulated combinatorial clustering as a QUBO and evaluated it on D-Wave hardware. Date et al.~\cite{Date2021} developed QUBO formulations for training machine learning models including balanced k-means clustering, and noted that encoding precision presents a fundamental trade-off between solution accuracy and QUBO problem size. Our work differs from these approaches in that we target the initialization stage of k-means rather than the full clustering assignment, and introduce an iterative refinement mechanism to recover real-valued centroids from binary outputs.

\subsection{QOCI: Quantum Optimized Centroid Initialization}

In prior work \cite{Allgood2023}, we introduced Quantum Optimized Centroid
Initialization (QOCI), which formulates centroid selection as a QUBO problem
derived from a matrix factorization framework. The data matrix is decomposed
as $V \approx WH$, where columns of $W$ encode candidate centroid coordinates
and $H$ is a binary assignment matrix. Minimizing $\|V - WH\|^2$ with respect
to $W$ and $H$ yields a quadratic objective in binary variables that can be
submitted to a quantum annealer or compatible solver. This formulation builds on \
the quantum annealing approach to linear least squares problems introduced by Borle and Lomonaco~\cite{Borle2019}, which established the QUBO encoding of real-valued variables using two's complement binary representation and analyzed the conditions under which quantum annealing may outperform classical methods for such problems.

QOCI demonstrated that centroid initialization could be cast as a QUBO and
solved on quantum annealing hardware, producing clusterings comparable to
random initialization. However, the method had several limitations: centroid
values were restricted to integers due to the binary encoding, the
formulation could not scale to larger sample sizes within D-Wave's free
account time constraints, and penalty parameters ($\delta_1$ for
linearization, $\delta_2$ for the assignment constraint on $H$) required
manual tuning. AQOCI, presented in Section~\ref{sec:contribution}, addresses
these limitations through an adaptive iterative refinement scheme.

\section{Contribution}\label{sec:contribution}

Adaptive Quantum Optimized Centroid Initialization (AQOCI) addresses the
principal limitations of QOCI: the restriction to integer-valued centroids,
limited scalability within quantum annealing time constraints, and the
absence of a mechanism for iterative refinement. The key insight is to
treat the QUBO solution not as a final answer but as the starting point
of an iterative process that progressively narrows the solution space,
recovering real-valued centroid coordinates from binary outputs through
adaptive scaling and offset parameters.

\subsection{Formulation}

As in QOCI, we begin from the matrix factorization $V \approx WH$, where
$V$ is the data matrix, $W$ encodes centroid coordinates, and $H$ is a
binary assignment matrix. The objective $\|V - WH\|^2$ is expanded into a
quadratic polynomial in binary variables, yielding linear and quadratic
coefficients that define a QUBO instance. AQOCI modifies this formulation
by introducing a scale parameter $\lambda$ and an offset parameter applied
to the polynomial expansion:
\begin{equation}
    \lambda\big(v_{11}^{2} + v_{12}^{2} + v_{21}^{2} + v_{22}^{2}
    - 2h_{11}v_{11}w_{11} \cdots \big) + \textit{offset}
\end{equation}
where $\lambda$ controls the resolution of the binary encoding and the
offset shifts the representable range to cover the region of interest in
the data space.

The binary encoding represents centroid coordinate values using $p$ qubits
plus a sign qubit. For a given set of qubits $\{q_{j\theta}\}$, the
encoded value is:
\begin{equation}
    \sum_{\theta \in \Theta} 2^{\theta} q_{j\theta}
\end{equation}
where $\Theta$ is determined by the number of available qubits. The sign
convention follows two's complement, with the most significant qubit
designating the sign.

\subsection{Adaptive Iterative Refinement}

The central contribution of AQOCI is an iterative refinement mechanism
inspired by the Gauss-Seidel and Jacobi methods for solving linear
systems \cite{Pollachini2021}. Rather than solving the full QUBO in a
single pass, AQOCI decomposes the problem into a sequence of smaller
QUBO instances, each refining the solution from the previous iteration.

At each iteration, a QUBO is formulated and submitted to the solver. The
returned binary string is interpreted as a real value using the current
scale and offset parameters, which are then updated to narrow the search
window around the current solution. The scale parameter is initialized as:
\begin{equation}
    s = \frac{\textit{upperLimit} - \textit{lowerLimit}}{2^{\textit{bits}} - 1}
\end{equation}
where the upper and lower limits define the representable range given
the available qubits. A scale list and offset list, each of length equal
to the number of substitution variables, track the per-variable resolution
and shift.

Within each iteration, the scale and offset are updated according to the
solver output. Let $\alpha$ denote the current scale value, $\beta$ the
scale factor (set to 2 in our experiments), $o$ the current offset, $b$
the integer value of the returned binary string, and $ll$ and $ul$ the
lower and upper limits respectively. The update rule is:
\begin{equation}
  \begin{cases}
    \alpha \leftarrow \alpha/\beta,\quad o \leftarrow o
      & \text{if } b = 0 \text{ and } o = ll \\
    \alpha \leftarrow \alpha/\beta,\quad o \leftarrow b\alpha + o - b\cdot o
      & \text{if } b = 2^{\textit{bits}}-1 \text{ and } ul = (2^{\textit{bits}}-1)\alpha + ll \\
    \alpha \leftarrow \alpha/\beta,\quad o \leftarrow b\alpha + o - 3\alpha
      & \text{otherwise}
  \end{cases}
\end{equation}

At each iteration the scale is halved, progressively increasing the
precision of the represented centroid values. The offset adjusts to
re-center the representable window around the current best estimate. This
process continues for a fixed number of iterations determined empirically;
while a tolerance-based stopping criterion of the form
$\varepsilon^k = \|x^{(k)} - x\| / \|x\| \leq t$ is standard for
iterative solvers, we found that the $L_2$ norm fluctuated across
iterations for larger QUBO instances, preventing reliable convergence
detection. A fixed iteration count provided more consistent behavior
across problem sizes.

\subsection{Computational Considerations}

The iterative decomposition serves a dual purpose. First, it enables
real-valued centroid recovery from inherently discrete binary outputs by
progressively refining the resolution. Second, it breaks the computation
into smaller blocks that fit within the time constraints of quantum
annealing platforms. At the time of this work, D-Wave provided limited
monthly computation time on free developer accounts, making single-pass
solutions to larger problem instances infeasible.

The QUBO formulation is solver-agnostic: the same instance can be
submitted to quantum annealers, hybrid quantum-classical solvers, or
purely classical heuristics without modification. In our experiments, we
evaluate three backends provided by D-Wave's Ocean software platform
\cite{dwave}: TABU search \cite{mst2}, simulated annealing, and the
Hybrid BQM solver, which combines classical and quantum processing
through a proprietary decomposition.

Once AQOCI produces candidate centroids, these are passed to standard
$k$-means as initial centers (via the \texttt{init} parameter in
scikit-learn), and $k$-means proceeds with its usual alternating
assignment and update steps until convergence. The quality of the
resulting clustering is then evaluated against the same metrics used
for all other initialization methods, ensuring a fair comparison.

\section{Experimental Setup}

\subsection{Datasets}

We evaluate AQOCI on two categories of data: synthetic Gaussian blobs with controlled parameters and the MOTIF malware classification dataset. 

\textbf{Synthetic Gaussian data.} To systematically characterize the conditions under which QUBO-based initialization offers advantages over classical methods, we conduct four experimental sweeps on synthetic data generated using scikit-learn's \texttt{make\_blobs}, each varying a single factor while holding all others fixed. In all synthetic experiments, $n$ denotes the number of data points, $k$ the number of clusters, $d$ the dimensionality of the feature space, and $\sigma$ the standard deviation of each Gaussian cluster. Each sweep varies a single parameter while holding the remaining three fixed:

\begin{enumerate}
    \item \textbf{Cluster separation sweep.} Fixed $k=3$, $d=2$, $n=250$, $\sigma=0.5$. We vary inter-cluster distance across five configurations: heavy overlap (center spacing $\approx 1.0$), moderate overlap (the configuration from our prior work~\cite{Allgood2023}, with centers at $[1,6]$, $[2,4]$, $[3,5]$), mild separation (spacing $\approx 7$), well-separated (spacing $\approx 14$--$20$), and extreme separation (spacing $\approx 50$--$100$). This sweep directly tests the hypothesis that AQOCI's global optimization approach provides greater benefit when cluster boundaries are ambiguous.

    \item \textbf{Cluster count ($k$) sweep.} Fixed $d=2$, $n=500$, $\sigma=0.5$. We vary $k \in \{3, 5, 7, 10\}$ with centers placed evenly on a circle of radius 10 to maintain consistent inter-cluster geometry. This tests how the QUBO formulation scales with the number of clusters, given that k-means++ has an $O(\log k)$ approximation guarantee.

    \item \textbf{Dimensionality sweep.} Fixed $k=3$, $n=500$, $\sigma=0.5$. We vary $d \in \{2, 5, 10\}$ with centers spaced at intervals of 10 along the first axis to isolate the effect of dimensionality from cluster separation. This addresses whether the QUBO formulation, which scales in variable count with $p \times k$, remains viable in higher-dimensional feature spaces.

    \item \textbf{Sample size sweep.} Fixed $k=3$, $d=2$, $\sigma=0.5$, with moderate separation (centers at $[0,0]$, $[5,5]$, $[10,0]$). We vary $n \in \{50, 100, 250, 500\}$. This tests scaling behavior with classical solvers alone, as quantum annealing hardware constraints limited our original experiments to $n \leq 250$.
\end{enumerate}

\textbf{MOTIF malware dataset.} The MOTIF dataset~\cite{Joyce2023} provides EMBER feature vectors for malware samples with ground-truth family labels. We normalize feature vectors using min-max scaling and reduce to two dimensions via Principal Component Analysis (PCA). Samples are drawn sequentially from the training set with random seed 0. Experiments are conducted at sample sizes $n \in \{20, 30, 35, 40, 45, 50, 100, 250\}$ with $k = 3$ clusters.

\subsection{Methods}

We compare four initialization strategies:

\begin{enumerate}
    \item \textbf{Random initialization.} Standard k-means with \texttt{init='random'}, selecting $k$ observations uniformly at random as initial centroids.
    \item \textbf{k-means++ initialization.} Standard k-means with \texttt{init='k-means++'}~\cite{Arthur2007}, selecting centroids with distance-weighted probability.
    \item \textbf{AQOCI--TABU.} Centroids computed via the AQOCI formulation solved with the TABU search heuristic~\cite{mst2}, then passed to k-means as initial centers.
    \item \textbf{AQOCI--SimAnn.} Centroids computed via AQOCI solved with simulated annealing, then passed to k-means.
    \item \textbf{AQOCI--Hybrid.} Centroids computed via AQOCI solved with D-Wave's Hybrid BQM solver~\cite{dwave}, which combines classical and quantum processing through a proprietary decomposition, then passed to k-means as initial centers. This solver was evaluated on the MOTIF dataset only, as D-Wave's free developer program was discontinued prior to the expanded synthetic experiments.
\end{enumerate}

In our prior work, we additionally evaluated D-Wave's Hybrid BQM solver, which incorporates quantum annealing hardware. Those results are retained for the MOTIF and original synthetic experiments. However, D-Wave's free developer program has since been discontinued, and the expanded synthetic experiments reported here use only the classical solver backends (TABU and simulated annealing). The QUBO formulation is solver-agnostic: the same instance can be submitted to any compatible backend without modification, and the original Hybrid results demonstrated that all three solvers converge to comparable performance at the problem sizes tested.

For all methods, k-means is run with \texttt{n\_init=10}, \texttt{max\_iter=10000}, and tolerance $10^{-4}$. For the AQOCI variants, solver parameters include \texttt{num\_reads=2000}, \texttt{num\_sweeps=2000}, and TABU timeout of 10 seconds. The number of adaptive iterations is scaled with problem size: 4 iterations for $n \leq 50$, 6 for $n \leq 100$, and 8 for larger instances.

\subsection{Evaluation Metrics}

We report six standard clustering metrics:

\begin{itemize}
    \item \textbf{Inertia}: within-cluster sum of squared distances (lower is better).
    \item \textbf{Iterations}: number of k-means iterations to convergence (lower is better).
    \item \textbf{Silhouette score}: measure of cluster cohesion and separation, ranging from $-1$ to $+1$ (higher is better).
    \item \textbf{Homogeneity}: whether each cluster contains only members of a single ground-truth class (higher is better).
    \item \textbf{Completeness}: whether all members of a ground-truth class are assigned to the same cluster (higher is better).
    \item \textbf{V-measure}: harmonic mean of homogeneity and completeness (higher is better).
\end{itemize}

\section{Results}

\subsection{Synthetic Gaussian Data}

\subsubsection{Cluster Separation}

The separation sweep reveals the central finding of our expanded experiments: AQOCI's relative performance depends directly on the degree of cluster overlap. Table~\ref{tab:separation} summarizes V-measure results across all five configurations.

\begin{table}[h]
\centering
\caption{V-measure on synthetic data across cluster separation levels ($n=250$, $k=3$, $d=2$, $\sigma=0.5$). Bold indicates best performance per configuration.}
\label{tab:separation}
\begin{tabular}{lcccc}
\hline
Configuration & Random & k-means++ & TABU & SimAnn \\
\hline
Heavy overlap   & 0.329 & 0.329 & 0.174 & \textbf{0.350} \\
Moderate overlap & \textbf{0.779} & \textbf{0.779} & 0.536 & 0.536 \\
Mild separation  & \textbf{1.000} & \textbf{1.000} & 0.648 & 0.648 \\
Well-separated   & \textbf{1.000} & \textbf{1.000} & 0.648 & 0.652 \\
Extreme separation & \textbf{1.000} & \textbf{1.000} & \textbf{1.000} & \textbf{1.000} \\
\hline
\end{tabular}
\end{table}

On heavily overlapping clusters, where center spacing is approximately $1.0$ with $\sigma = 0.5$, AQOCI--SimAnn achieves the highest V-measure of 0.350, exceeding both k-means++ and random initialization (0.329). This is consistent with the MOTIF results reported in Section~\ref{sec:motif}, where AQOCI similarly outperforms k-means++ on data with complex, overlapping structure. AQOCI--TABU performs poorly on this configuration (0.174), suggesting that solver choice affects solution quality when cluster boundaries are ambiguous.

From moderate overlap onward, k-means++ achieves perfect or near-perfect clustering and AQOCI falls behind substantially. At mild and well-separated configurations, both AQOCI variants plateau at a V-measure of approximately 0.648 despite inertia values an order of magnitude higher than k-means++ (e.g., 2204 vs.\ 123 at mild separation). At extreme separation, where clusters are unambiguous regardless of initialization strategy, all methods converge to perfect V-measure of 1.0.

The consistent plateau at $V \approx 0.648$ for AQOCI across the mild, well-separated, and extremely separated configurations indicates a systematic limitation rather than random failure. We attribute this to the binary encoding resolution: with 3 bits per variable, the representable range is discretized into $2^3 - 1 = 7$ intervals. When the data range is large relative to the cluster spacing, the adaptive refinement cannot place centroids with sufficient precision to distinguish nearby clusters, regardless of the number of iterations. When clusters are compact and overlapping, the same 3-bit resolution spans a narrower range and can capture finer distinctions in centroid placement.

\subsubsection{Cluster Count}

Table~\ref{tab:ksweep} shows V-measure results as the number of clusters increases from 3 to 10, with centers placed evenly on a circle of radius 10.

\begin{table}[h]
\centering
\caption{V-measure across cluster counts ($n=500$, $d=2$, $\sigma=0.5$, centers on circle of radius 10). TABU failed at $k=10$ due to a variable naming collision in the QUBO matrix construction. Bold indicates best performance per $k$.}
\label{tab:ksweep}
\begin{tabular}{lcccc}
\hline
$k$ & Random & k-means++ & TABU & SimAnn \\
\hline
3  & \textbf{1.000} & \textbf{1.000} & \textbf{1.000} & \textbf{1.000} \\
5  & \textbf{1.000} & \textbf{1.000} & 0.865 & 0.866 \\
7  & \textbf{1.000} & \textbf{1.000} & \textbf{1.000} & 0.922 \\
10 & \textbf{1.000} & \textbf{1.000} & --- &  0.914 \\
\hline
\end{tabular}
\end{table}

At $k=3$, AQOCI matches k-means++ with perfect clustering. As $k$ increases, AQOCI performance degrades while k-means++ remains perfect. At $k=5$, both AQOCI variants produce V-measures near 0.865. At $k=7$, AQOCI--TABU recovers to perfect performance while AQOCI--SimAnn achieves 0.922. At $k=10$, AQOCI--TABU fails due to an implementation-level variable naming collision in the QUBO matrix that produces an asymmetric $Q$ (discussed in Section~\ref{sec:limitations}), while AQOCI--SimAnn achieves 0.914.

Despite the degradation, the V-measures remain above 0.86 even at $k=10$, indicating that k-means is able to partially recover from imprecise AQOCI initializations through its iterative refinement. The inconsistent behavior across solvers (TABU outperforming SimAnn at $k=7$ but failing at $k=10$) reflects the stochastic nature of heuristic solvers operating on increasingly large QUBO instances.

\subsubsection{Dimensionality}

Table~\ref{tab:dimensionality} shows results as the feature dimensionality increases from 2 to 10, with cluster centers spaced at intervals of 10 along the first axis.

\begin{table}[h]
\centering
\caption{V-measure across feature dimensions ($n=500$, $k=3$, $\sigma=0.5$, center spacing 10 along axis 0). TABU failed at $d=10$ due to the same variable naming issue as $k=10$. Bold indicates best performance per $d$.}
\label{tab:dimensionality}
\begin{tabular}{lcccc}
\hline
$d$ & Random & k-means++ & TABU & SimAnn \\
\hline
2  & \textbf{1.000} & \textbf{1.000} & 0.649 & 0.648 \\
5  & \textbf{1.000} & \textbf{1.000} & 0.649 & 0.648 \\
10 & \textbf{1.000} & \textbf{1.000} & --- & \textbf{1.000} \\
\hline
\end{tabular}
\end{table}

The most notable result is at $d=10$: AQOCI--SimAnn matches k-means++ with a perfect V-measure of 1.0 and identical inertia (1206.59). This demonstrates that the QUBO formulation scales to higher-dimensional feature spaces without inherent degradation. The number of $W$ variables grows linearly with $d$ (specifically $d \times k$), increasing the QUBO size, but the formulation remains tractable for classical solvers at the dimensions tested.

At $d=2$ and $d=5$, AQOCI produces the same $V \approx 0.648$ plateau observed in the separation sweep. Since the cluster separation in these configurations (spacing of 10 with $\sigma=0.5$) falls in the well-separated regime, the degradation is attributable to the encoding resolution limitation rather than dimensionality itself. The $d=10$ result, where the same cluster separation yields perfect performance, may reflect the higher-dimensional QUBO landscape providing the solver with more degrees of freedom to find accurate centroid placements.

\subsubsection{Sample Size}

Table~\ref{tab:samplesize} shows results as the number of data points increases from 50 to 500, using moderate cluster separation.

\begin{table}[h]
\centering
\caption{V-measure across sample sizes ($k=3$, $d=2$, $\sigma=0.5$, moderate separation). Bold indicates best performance per $n$.}
\label{tab:samplesize}
\begin{tabular}{lcccc}
\hline
$n$ & Random & k-means++ & TABU & SimAnn \\
\hline
50  & \textbf{1.000} & \textbf{1.000} & 0.688 & \textbf{1.000} \\
100 & \textbf{1.000} & \textbf{1.000} & 0.650 & 0.650 \\
250 & \textbf{1.000} & \textbf{1.000} & 0.648 & 0.648 \\
500 & \textbf{1.000} & \textbf{1.000} & 0.648 & 0.648 \\
\hline
\end{tabular}
\end{table}

At $n=50$, AQOCI--SimAnn matches k-means++ with perfect clustering, demonstrating that the QUBO formulation can find optimal initializations at small sample sizes when the solver succeeds. AQOCI--TABU achieves a lower V-measure of 0.688, indicating solver-dependent variability at small $n$. From $n=100$ onward, both AQOCI variants converge to the same $V \approx 0.648$ plateau observed across the other sweeps, while k-means++ maintains perfect performance.

The sample size sweep uses moderate separation (center spacing $\approx 7$), which falls in the regime where k-means++ dominates. The convergence of both AQOCI variants to the same plateau regardless of $n$ confirms that the limiting factor is encoding resolution rather than sample size. This contrasts with the MOTIF results (Section~\ref{sec:motif}), where AQOCI shows stronger performance at small $n$ on data with more complex structure, suggesting that the interaction between sample size and data complexity determines when QUBO-based initialization is most beneficial.

\subsection{MOTIF Malware Data}\label{sec:motif}

The MOTIF results present a markedly different picture. At smaller
sample sizes, AQOCI, particularly with the TABU solver, produces
initializations that lead to higher-quality clusterings than both
random initialization and $k$-means++. At $n=20$, AQOCI--TABU achieves
a V-measure of $0.552$, exceeding $k$-means++ ($0.530$) and random
initialization ($0.461$). At $n=45$, AQOCI--TABU reaches $0.452$
compared to $0.359$ for both random and $k$-means++, a relative
improvement of approximately $26\%$.

At $n=50$, both AQOCI--TABU and AQOCI--Hybrid achieve $0.435$, again
outperforming $k$-means++ ($0.391$) and random initialization ($0.401$).
At $n=100$, AQOCI--TABU retains a modest advantage ($0.308$ vs.\
$0.281$). By $n=250$, all methods converge to the same V-measure of
$0.237$, indicating that at larger sample sizes the data structure
dominates and initialization strategy has diminishing impact.

The completeness scores across AQOCI variants on MOTIF are notably
high, frequently reaching $1.0$ at smaller sample sizes, indicating
that the QUBO-derived initializations tend to place all members of a
ground-truth class into the same cluster. Homogeneity is lower,
reflecting the tendency to merge distinct classes into single clusters.
This asymmetry suggests that the AQOCI formulation captures global
structure in the data effectively but at a resolution too coarse to
separate nearby classes which is a limitation consistent with the binary
encoding precision.

\begin{table}[htpb]
\centering
\caption{Summary of clustering metrics on MOTIF malware data at selected
sample sizes. Bold values indicate best performance per metric and
sample size.}
\label{tab:motif_summary}
\small
\begin{tabular}{llccccc}
\toprule
\textbf{Metric} & \textbf{$n$} & \textbf{Random} & \textbf{TABU} &
\textbf{SimAnn} & \textbf{Hybrid} & \textbf{$k$-means++} \\
\midrule
\multirow{3}{*}{Inertia $\downarrow$}
  & 20  & \textbf{18.12} & 21.44 & 19.79 & 18.34 & 18.18 \\
  & 50  & 43.25 & 49.98 & 43.38 & 49.98 & \textbf{43.21} \\
  & 250 & 242.74 & 242.73 & 242.74 & 242.74 & 242.74 \\
\midrule
\multirow{3}{*}{Iterations $\downarrow$}
  & 20  & 4  & 4  & \textbf{3}  & 4  & \textbf{2} \\
  & 50  & 8  & 5  & 9  & 5  & \textbf{4} \\
  & 250 & 15 & 9  & 11 & 14 & \textbf{6} \\
\midrule
\multirow{3}{*}{Silhouette $\uparrow$}
  & 20  & 0.413 & 0.357 & 0.391 & \textbf{0.427} & 0.425 \\
  & 50  & 0.493 & 0.389 & \textbf{0.509} & 0.389 & 0.507 \\
  & 250 & \textbf{0.528} & \textbf{0.528} & \textbf{0.528} & \textbf{0.528} & \textbf{0.528} \\
\midrule
\multirow{3}{*}{Homogeneity $\uparrow$}
  & 20  & 0.299 & \textbf{0.381} & 0.300 & 0.296 & 0.360 \\
  & 50  & 0.251 & \textbf{0.280} & 0.234 & \textbf{0.280} & 0.243 \\
  & 250 & 0.140 & 0.140 & 0.140 & 0.140 & 0.140 \\
\midrule
\multirow{3}{*}{Completeness $\uparrow$}
  & 20  & \textbf{1.000} & \textbf{1.000} & 0.925 & \textbf{1.000} & \textbf{1.000} \\
  & 50  & \textbf{1.000} & 0.974 & \textbf{1.000} & 0.974 & \textbf{1.000} \\
  & 250 & 0.768 & 0.768 & 0.768 & 0.768 & 0.768 \\
\midrule
\multirow{3}{*}{V-measure $\uparrow$}
  & 20  & 0.461 & \textbf{0.552} & 0.453 & 0.457 & 0.530 \\
  & 50  & 0.401 & \textbf{0.435} & 0.379 & \textbf{0.435} & 0.391 \\
  & 250 & 0.237 & 0.237 & 0.237 & 0.237 & 0.237 \\
\bottomrule
\end{tabular}
\end{table}

\begin{figure}[!t]
\centering
\begin{subfigure}[t]{0.485\textwidth}
\centering
\begin{tikzpicture}
\begin{axis}[
    width=\linewidth,
    height=6.5cm,
    xlabel={Samples},
    ylabel={V-Measure},
    xmin=15, xmax=260,
    ymin=0.15, ymax=0.60,
    xtick={20,50,100,150,200,250},
    grid=major,
    grid style={gray!30},
    tick label style={font=\small},
    label style={font=\small},
    title={\small (a) V-Measure},
]
\addplot[color=clrRandom, mark=square*, thick, mark size=2pt]
    coordinates {(20,0.461)(30,0.426)(35,0.392)(40,0.374)
                 (45,0.359)(50,0.401)(100,0.281)(250,0.237)};
\addplot[color=clrTabu, mark=triangle*, thick, mark size=2pt]
    coordinates {(20,0.552)(30,0.463)(35,0.448)(40,0.374)
                 (45,0.452)(50,0.435)(100,0.308)(250,0.237)};
\addplot[color=clrSim, mark=diamond*, thick, mark size=2pt]
    coordinates {(20,0.453)(30,0.426)(35,0.392)(40,0.374)
                 (45,0.359)(50,0.379)(100,0.281)(250,0.237)};
\addplot[color=clrHybrid, mark=pentagon*, thick, mark size=2pt]
    coordinates {(20,0.457)(30,0.431)(35,0.432)(40,0.415)
                 (45,0.359)(50,0.435)(100,0.281)(250,0.237)};
\addplot[color=clrKpp, mark=o, thick, dashed, mark size=2pt]
    coordinates {(20,0.530)(30,0.426)(35,0.392)(40,0.343)
                 (45,0.359)(50,0.391)(100,0.281)(250,0.237)};
\end{axis}
\end{tikzpicture}
\end{subfigure}
\hfill
\begin{subfigure}[t]{0.485\textwidth}
\centering
\begin{tikzpicture}
\begin{axis}[
    width=\linewidth,
    height=6.5cm,
    xlabel={Samples},
    ylabel={Silhouette Score},
    xmin=15, xmax=260,
    ymin=0.30, ymax=0.56,
    xtick={20,50,100,150,200,250},
    grid=major,
    grid style={gray!30},
    tick label style={font=\small},
    label style={font=\small},
    title={\small (b) Silhouette Score},
]
\addplot[color=clrRandom, mark=square*, thick, mark size=2pt]
    coordinates {(20,0.413)(30,0.510)(35,0.525)(40,0.519)
                 (45,0.527)(50,0.493)(100,0.518)(250,0.528)};
\addplot[color=clrTabu, mark=triangle*, thick, mark size=2pt]
    coordinates {(20,0.357)(30,0.463)(35,0.483)(40,0.519)
                 (45,0.379)(50,0.389)(100,0.501)(250,0.528)};
\addplot[color=clrSim, mark=diamond*, thick, mark size=2pt]
    coordinates {(20,0.391)(30,0.510)(35,0.525)(40,0.519)
                 (45,0.527)(50,0.509)(100,0.518)(250,0.528)};
\addplot[color=clrHybrid, mark=pentagon*, thick, mark size=2pt]
    coordinates {(20,0.427)(30,0.490)(35,0.489)(40,0.467)
                 (45,0.527)(50,0.389)(100,0.518)(250,0.528)};
\addplot[color=clrKpp, mark=o, thick, dashed, mark size=2pt]
    coordinates {(20,0.425)(30,0.510)(35,0.525)(40,0.528)
                 (45,0.527)(50,0.507)(100,0.518)(250,0.528)};
\end{axis}
\end{tikzpicture}
\end{subfigure}

\vspace{2pt}
\centering
\begin{tikzpicture}
\begin{axis}[
    hide axis,
    xmin=0, xmax=1, ymin=0, ymax=1,
    legend columns=5,
    legend style={draw=gray!50, font=\small, column sep=6pt,
                  anchor=center, at={(0.5,0.5)}},
    scale only axis, width=0.01cm, height=0.01cm,
]
\addlegendimage{color=clrRandom, mark=square*, thick, mark size=2pt}
\addlegendentry{Random}
\addlegendimage{color=clrTabu, mark=triangle*, thick, mark size=2pt}
\addlegendentry{TABU}
\addlegendimage{color=clrSim, mark=diamond*, thick, mark size=2pt}
\addlegendentry{SimAnn}
\addlegendimage{color=clrHybrid, mark=pentagon*, thick, mark size=2pt}
\addlegendentry{Hybrid}
\addlegendimage{color=clrKpp, mark=o, thick, dashed, mark size=2pt}
\addlegendentry{$k$-means++}
\end{axis}
\end{tikzpicture}

\caption{Clustering performance on MOTIF malware data. (a) V-measure
shows AQOCI--TABU achieving the highest scores at most sample sizes,
outperforming $k$-means++ by up to 26\% at $n=45$. (b) Silhouette
scores are comparable across all methods with convergence at larger
sample sizes.}
\label{fig:motif}
\end{figure}
\FloatBarrier

\subsection{Solver Comparison}

Across both datasets, the three AQOCI solver backends produce
broadly similar results, with the TABU solver showing a slight
advantage on the MOTIF data. At $n=250$, all three solvers converge
to essentially identical clusterings on both datasets, with
differences in inertia below $0.01$. This convergence suggests that
the QUBO formulation itself, rather than the choice of solver, is the
primary determinant of clustering quality at scale.

The Hybrid BQM solver, which incorporates quantum annealing hardware,
does not consistently outperform the purely classical TABU and
simulated annealing backends in our experiments. This is consistent
with the small problem sizes tested: at $n=250$ with $k=3$ in two
dimensions, the QUBO instances are small enough that classical
heuristics find near-optimal solutions efficiently. Larger problem
instances, where classical solvers face combinatorial scaling, may
reveal differences between solver backends that are not visible at
this scale.

\subsection{Discussion}

The expanded synthetic experiments, combined with the MOTIF results, reveal two factors that govern AQOCI's relative performance: the structure of the data and the precision of the binary encoding.

\textbf{Data structure.} AQOCI's global optimization approach provides the greatest benefit when cluster structure is complex or overlapping. On heavily overlapping synthetic clusters, AQOCI--SimAnn outperforms k-means++ (V-measure 0.350 vs.\ 0.329). On the MOTIF malware dataset, which exhibits complex, non-Gaussian structure after PCA reduction, AQOCI--TABU outperforms k-means++ by up to 26\% at small sample sizes. In both cases, the QUBO formulation's global perspective over the full data matrix identifies initializations that the greedy, distance-based k-means++ misses. When cluster structure is simple and well-separated, local decisions suffice and k-means++ is clearly superior.

\textbf{Encoding resolution.} The consistent V-measure plateau at approximately 0.648 across the separation, dimensionality, and sample size sweeps points to a systematic bottleneck in the binary encoding. With $p = 3$ qubits per variable, each coordinate is discretized into $2^p - 1 = 7$ representable intervals within the adaptive refinement window. When the data range is large relative to the inter-cluster distance, this resolution is insufficient to place centroids with the precision needed to separate clusters. When the data is compact --- as in the heavy overlap configuration or the MOTIF dataset after normalization and PCA --- the same 3-bit encoding spans a narrower range and achieves finer effective precision. Ramamurthy et al.~\cite{Ramamurthy2024} analyze a related phenomenon in the context of solving linear systems via QUBO, showing that the standard box contraction ratio of 0.5, equivalent to our scale factor $\beta = 2$, is suboptimal and that a ratio of 0.2 achieves faster convergence. Adapting such optimized contraction strategies to the AQOCI refinement loop is a promising direction for improving centroid precision without increasing the number of qubits.

This encoding resolution limitation is not fundamental to the QUBO formulation but rather a parameter of the current implementation. Increasing the number of bits per variable would improve precision at the cost of a larger QUBO matrix, creating a direct trade-off between centroid accuracy and computational tractability that future work could characterize systematically.

\textbf{Solver behavior.} Across the synthetic experiments, AQOCI--SimAnn generally produces more consistent results than AQOCI--TABU, particularly at smaller sample sizes and higher cluster counts. At $n=50$, SimAnn achieves perfect clustering while TABU does not. At $k=7$, the pattern reverses, with TABU achieving perfect clustering while SimAnn falls short. This variability reflects the stochastic nature of heuristic solvers on QUBO landscapes that grow in complexity with problem size, and suggests that solver selection or ensembling may offer further improvements.

\textbf{Scalability.} The dimensionality sweep demonstrates that the QUBO formulation scales to at least $d=10$ without inherent degradation in clustering quality: AQOCI--SimAnn achieves perfect V-measure at $d=10$, matching k-means++. The TABU solver encounters an implementation-level failure at $d=10$ and $k=10$ (discussed in Section~\ref{sec:limitations}), but this is a variable naming issue in the current codebase rather than a limitation of the formulation. The sample size sweep confirms tractability up to $n=500$ with classical solvers, though runtime increases substantially at larger $n$ due to the growth of the QUBO matrix.

\section{Limitations}\label{sec:limitations}

Several limitations of this work should be noted, spanning both the QUBO formulation and the current implementation.

\textbf{Encoding resolution.} The expanded synthetic experiments reveal a consistent V-measure plateau at approximately 0.648 when AQOCI is applied to data with well-separated clusters spanning a wide range. With 3 bits per variable, the adaptive refinement discretizes each coordinate into 7 representable intervals per iteration. While the iterative scheme progressively narrows this window, the precision achieved is insufficient to distinguish clusters when the data range is large relative to inter-cluster spacing. Increasing the number of bits per variable would improve precision but proportionally increases the QUBO matrix size, creating a trade-off between centroid accuracy and computational tractability that has not been systematically characterized.

\textbf{Penalty parameter sensitivity.} The penalty parameters $\delta_1$ (linearization) and $\delta_2$ (assignment constraint on $H$) were set to $\delta_1 = 50$ and $\delta_2 = 100$ across all experiments. At larger problem sizes, particularly at higher $k$ and $n$, the number of violated columns in the $H$ matrix increases substantially (e.g., 134 out of 500 at $k=3$, $n=500$; 435 out of 500 at $k=5$), indicating that these penalty values do not scale adequately with problem size. A systematic study of penalty parameter scaling, or an adaptive penalty scheme, would likely improve performance on larger instances.

\textbf{Implementation constraints.} The current variable naming scheme in the QUBO matrix construction uses concatenated indices (e.g., \texttt{w110} for row 1, column 10 of $W$), which creates ambiguity at $k \geq 10$ or $d \geq 10$ (e.g., \texttt{w110} could denote row 1 column 10 or row 11 column 0). This produces asymmetric QUBO matrices that cause the TABU solver to fail, though the simulated annealing solver is more tolerant. This is an engineering issue in the current codebase, not a limitation of the formulation, and would be resolved by adopting a delimited naming convention. Additionally, the QUBO matrix size at $d = 20$ exceeded available system memory, preventing evaluation at higher dimensionalities.

\textbf{Convergence behavior.} The adaptive refinement procedure uses a fixed number of iterations rather than a principled convergence criterion. The $L_2$ norm of the residual exhibits non-monotonic behavior across iterations, with norms frequently increasing rather than decreasing (e.g., from 18{,}074 to 27{,}261 over 8 iterations at $k=3$, $n=500$). This oscillation prevents the use of a tolerance-based stopping condition and suggests that the QUBO energy landscape contains local minima that the solvers traverse without consistent improvement. A more robust termination criterion, potentially based on monitoring the QUBO objective value or maintaining the best solution across iterations rather than the final one, would strengthen the method.

\textbf{Scale and scope.} The maximum sample size evaluated with AQOCI solvers is $n = 500$ for the expanded synthetic experiments and $n = 250$ for the MOTIF dataset, the latter constrained by the computational limits of D-Wave's free developer account at the time of the original experiments. This account has since been discontinued, precluding further evaluation with the Hybrid BQM solver. The experiments are conducted in two dimensions (or after PCA reduction to two dimensions for MOTIF), with the dimensionality sweep extending to $d = 10$. The behavior of AQOCI on higher-dimensional data without dimensionality reduction remains an open question. Evaluation on a broader range of real-world datasets would provide a more comprehensive picture of the conditions under which QUBO-based initialization offers practical advantages.

\section{Conclusion}

We have presented Adaptive Quantum Optimized Centroid Initialization (AQOCI), a method that formulates the centroid initialization problem for prototype-based clustering as a QUBO instance solvable by quantum annealing or quantum-inspired classical solvers. AQOCI extends the prior QOCI algorithm by introducing an iterative refinement scheme that recovers real-valued centroid coordinates from binary solver outputs through adaptive scaling and offset updates.

Our experimental evaluation encompasses both synthetic Gaussian data with controlled sweeps over cluster separation, cluster count, dimensionality, and sample size and the MOTIF malware classification dataset, comparing against random initialization and k-means++. The results identify two factors that govern AQOCI's relative performance: the structure of the underlying data and the precision of the binary encoding.

On data with complex or overlapping cluster structure, AQOCI produces competitive and sometimes superior initializations. On the MOTIF malware dataset, AQOCI--TABU achieves V-measure improvements of up to 26\% over k-means++ at small sample sizes. On synthetic data with heavy cluster overlap, AQOCI--SimAnn outperforms k-means++ in V-measure (0.350 vs.\ 0.329). On well-separated synthetic clusters, k-means++ is clearly superior, and AQOCI exhibits a consistent V-measure plateau at approximately 0.648 attributable to the 3-bit encoding resolution. All methods converge to equivalent performance at larger sample sizes on the MOTIF data, indicating that initialization strategy has the greatest impact in small-sample regimes.

The dimensionality sweep demonstrates that the QUBO formulation scales to at least $d = 10$ without inherent degradation, with AQOCI--SimAnn matching k-means++ at perfect V-measure. The cluster count sweep shows graceful degradation up to $k = 10$, with V-measures remaining above 0.86 even when AQOCI centroids are imprecise, as k-means iterative refinement partially compensates for initialization error.

These results establish QUBO-based centroid initialization as a viable alternative to classical heuristics whose relative advantage is data-structure-dependent and encoding-precision-dependent. The formulation is solver-agnostic and can be executed on quantum annealing hardware, hybrid quantum-classical platforms, or purely classical solvers without modification. Several concrete directions for improvement follow from our findings: increasing the number of bits per variable to improve centroid precision, developing adaptive penalty parameter scaling to maintain $H$-matrix constraint satisfaction at larger problem sizes, and adopting a best-solution-across-iterations strategy to mitigate the observed non-monotonic convergence behavior. As quantum annealing hardware matures with increases in qubit count, connectivity, and coherence times, the potential for this formulation to deliver quantum speedup on larger, higher-dimensional clustering problems remains an open and promising direction.

\subsection*{Future Work}

Several directions for future investigation follow from this work.
Extending AQOCI to higher-dimensional data without PCA reduction
would test the formulation's scalability and practical applicability.
Developing a principled convergence criterion for the adaptive
iteration, potentially based on monitoring the QUBO objective value
rather than the centroid residual, could improve reliability and
reduce the need for manual parameter specification. Evaluating AQOCI
on a broader range of datasets and clustering tasks would provide a
more comprehensive picture of the conditions under which QUBO-based
initialization offers advantages over classical methods. Finally,
benchmarking on next-generation quantum annealing hardware with
larger qubit counts would directly address the question of whether
the formulation can realize a practical quantum advantage for
clustering at scale.
%\begin{appendices}

%\section{Section title of first appendix}\label{secA1}

%\end{appendices}

%^\bibliographystyle{sn-basic}
%\bibliographystyle{unsrt}
%\bibliography{bibliography}
\bibliography{bibliography}


\end{document}